\documentclass[twocolumn,aps,prx,floatfix,superscriptaddress,nofootinbib]{revtex4-2}
\usepackage{amsmath}
\usepackage{amsfonts}
\usepackage{amssymb}
\usepackage{graphicx}
\usepackage{color}
\usepackage{bm}
\usepackage{dsfont}
\usepackage{soul}
\usepackage{mathtools}

\DeclarePairedDelimiterX\braket[2]{\langle}{\rangle}{#1 \delimsize\vert #2}
\usepackage[final]{hyperref} 
\hypersetup{
	colorlinks=true,       
	linkcolor=blue,        
	citecolor=blue,        
	filecolor=magenta,     
	urlcolor=blue         
}

\newcommand{\new}[1]{\textcolor{black}{#1}}
\newcommand{\beq}{\begin{equation}}
\newcommand{\eneq}{\end{equation}}

\begin{document}

\title{Speeding up Pontus-Mpemba effects via dynamical phase transitions}

\author{Andrea Nava}
\affiliation{Institut f\"ur Theoretische Physik, Heinrich-Heine-Universit\"at, 40225 D\"usseldorf, Germany}
\author{Reinhold Egger}
\affiliation{Institut f\"ur Theoretische Physik, Heinrich-Heine-Universit\"at, 40225 D\"usseldorf, Germany}
\author{Bidyut Dey}
\affiliation{I.N.F.N., Gruppo collegato di Cosenza, 
Arcavacata di Rende I-87036, Cosenza, Italy}
\author{Domenico Giuliano}
\affiliation{Institut f\"ur Theoretische Physik, Heinrich-Heine-Universit\"at, 40225 D\"usseldorf, Germany}
\affiliation{I.N.F.N., Gruppo collegato di Cosenza, 
Arcavacata di Rende I-87036, Cosenza, Italy}
\affiliation{Dipartimento di Fisica, Universit\`a della Calabria Arcavacata di 
Rende I-87036, Cosenza, Italy}

\begin{abstract}
We demonstrate that open quantum systems exhibiting dynamical phase transitions (DPTs) allow for  efficient protocols implementing the Pontus-Mpemba effect. The relaxation speed-up toward a predesignated target state is tied to the existence of a long metastable time window preceding the DPT and can be exploited in applications to systematically optimize quantum protocols. As paradigmatic example for the connection between DPTs and quantum Mpemba effects, we study one-dimensional (1D) interacting lattice fermions corresponding to a dissipative variant of the Gross-Neveu (GN) model. 
 \end{abstract}
\maketitle
 
\section{Introduction}\label{sec1}

Classical \cite{Mpemba1969,Lu2017,Lasanta2017,Klich2019,Chetrite2021,Ibanez2024}
and quantum \cite{Nava2019,Carollo2021,Rylands2024,Murciano2024,Moroder2024,Nava2024m,Wang2024,Chatterjee2023,Chatterjee2024,Liu2024,Liu2025b,Yu2025_symmetry} Mpemba effects are counterintuitive anomalous nonequilibrium relaxation phenomena which may occur after a rapid parameter quench. They have recently garnered a lot of attention  \cite{Ares2025,Teza2025}, mainly motivated by the quest for understanding the underlying physical mechanisms as well as by the promise of useful applications, e.g., for speeding up relaxation processes or for optimizing quantum protocols such as state preparation and cooling schemes. For example, comparing two thermal states with initial temperatures $T_{c}$ (cold) and
$T_{h}$ (hot), respectively, the Mpemba effect occurs if after a sudden quench of the initial temperature to the final temperature $T_{\rm eq}<T_{c}<T_{h}$, the initially hotter system
relaxes faster to the final equilibrium state than the colder one \cite{Mpemba1969}.  
In generalized protocols, in particular for the quantum case, temperature may be replaced by  other control parameters.  Recently, a modified (classical or quantum) protocol dubbed Pontus-Mpemba effect (PME)  has been proposed \cite{Nava2025,Yu2025}, see also Ref.~\cite{Gal2020},
where both system copies start from the \emph{same} initial state {\bf S} in control parameter space. The first copy now undergoes a parametric quench driving it toward a target state {\bf F} in a time span $t_{\rm SF}$.  The second copy instead will first be quenched toward a different final state {\bf A}, which would be reached after a time $t_{\rm SA}$. 
However, upon reaching an intermediate state {\bf I} at time $t_{\rm SI}<t_{\rm SA}$, the system is decoupled from this environment and, by a second parameter quench,  
 driven to the desired target state {\bf F} in a time $t_{\rm IF}$.  By definition, the PME takes place if the time for the two-step protocol is shorter than for the direct process, $t_{\rm SI} + t_{\rm IF} < t_{\rm SF}$ \cite{Nava2025}.   Conceptually, the PME protocol offers several advantages \cite{Nava2025} over standard single-step Mpemba protocols \cite{Ares2025,Teza2025}. In particular, the notion of a parameter distance becomes obsolete, the state {\bf I} can be an arbitrary non-thermal nonequilibrium state, and the time cost for heating up the second copy is directly taken into account. For given initial (${\bf S}$) and final (${\bf F}$) states,  {\bf I} and {\bf A} can be chosen in order to optimize the PME efficiency.   

A seemingly unrelated major recent development in nonequilibrium statistical physics concerns the study of DPTs \cite{Zvyagin2016,Heyl2018,Heyl2019}, where a parametric quench drives a quantum system across a phase boundary at a critical time $t_\ast$ after the quench.
At the time $t_*$ corresponding to the DPT,  matrix elements of the time evolution operator typically exhibit singular behavior.  
Studies of DPTs have given valuable information about the critical dynamics of closed quantum systems prepared in pure  
\cite{Heyl2013,Jurcevic2017,Yuzbashyan2006,Prufer2018,Yamamoto2021,Mazza2017,Rossini2020,Pellissetto2020} or mixed states \cite{Abeling2016,Bhattacharya2017,Lang2018}. Interestingly, DPTs also appear in open quantum systems coupled to environments (e.g., thermal baths) \cite{Dimeglio2020,Nava2024,Nava2023_s}. The quench must then connect two ordered phases with order parameters of different symmetry in order to realize a DPT rather than a conventional relaxation crossover. As function of time, the order parameter here slowly rearranges itself by evolving through a long ``metastable'' time window ${\cal M}$ before the DPT occurs at time $t_*$.  

In this \new{paper}, we uncover an intimate connection between DPTs and quantum Mpemba effects for open quantum systems, and show how this connection allows one to implement efficient PME protocols.  We illustrate this connection for a 1D correlated lattice fermion model realizing the GN model \cite{Gross1974}, including a finite coupling of the fermions to an environment. The quantum dynamics of this open system is studied through the Lindblad master equation approach \cite{Lindblad1976,Breuer2007}, using  a
time-dependent self-consistent mean field (SCMF) approximation 
\cite{Peronaci2015,Guimaraes2016,Nava2019,Nava2021,Nava2023,Nava2023_s,Nava2024,Cinnirella2024,Cinnirella2025}. \new{For 2D superconductor models, it has been shown \cite{Nava2023_s,Nava2024} that this approach recovers the results of standard self-consistency relations but also captures important correlations on top. We here have generalized this numerical method in order to allow for arbitrary spatial order parameter profiles, where a finite system-reservoir coupling strength ensures stable convergence to the steady state.  
Given the generality of the arguments below, we speculate that the mechanism 
put forward here applies to generic open quantum systems with DPTs,} independent of the specific model and/or approximations made in computing the dynamics. (For a related discussion of closed quantum systems, see  Ref.~\cite{Parez2025}.) In particular, we show if, and how, the metastable region ${\cal M}$ preceding the DPT allows one to drastically speed up the system relaxation dynamics under PME protocols. 

\new{We note that the conceptual link between metastability, relaxation anomalies, and Mpemba-like speedups has been established before in the realm of classical physics \cite{Lasanta2017,Lu2017,Klich2019,Chetrite2021}. In these frameworks, metastable regimes preceding slow crossovers can already explain accelerated relaxation when the system trajectory transiently bypasses or exploits these regions.  One may therefore view the phenomena described below (metastable regions preceding a DPT can speed up the PME) as a quantum generalization of earlier works on classical systems. Specifically, in our case, the singularities of the time evolution operator at the DPT time $t_*$, with a preceding metastable time window ${\cal M}$, replace the spectral gaps of a classical Markovian generator.  However, since DPTs are not the only possible reason for metastable time regions in quantum systems, the general mechanism described below could also appear in other scenarios involving metastability.
Since the metastable region ${\cal M}$ preceding the DPT is the key ingredient for the connection to Mpemba effects, critical exponents or scaling features related to the DPT are only of secondary importance.
It is also worth noting that apart from the GN model studied below,  essentially the same DPT-PME physics occurs in the 2D superconductor models studied in Refs.~\cite{Nava2023_s,Nava2024}. In addition,
let us emphasize that the SCMF approximation becomes exact for an $N$-flavor generalization of the GN model in the limit of large $N$, see also App.~\ref{appA}. In that case, the GN lattice model becomes equivalent to a model of $N$ coupled lattice fermion chains, which, for $N\to\infty$, can be viewed as a 2D system.  The DPT-PME interplay discussed below should therefore also take place in 2D systems.}

Let us first consider a case where the target state {\bf F} is in a disordered phase (zero order parameter) while the initial state {\bf S} is in an ordered phase.  A first quench now takes the system toward an auxiliary state {\bf A} within a different ordered phase, where one must pass through a DPT and thus encounters the long metastable time region ${\cal M}$. A second quench then drives the system from an intermediate state {\bf I} (along the trajectory ${\bf S}\to {\bf A}$) toward {\bf F} in a very short time since it requires the melting of a nonzero order parameter.  Despite the fact that the direct crossover ${\bf S} \to  {\bf F}$ does not involve a DPT and, therefore, no slowing down due to ${\cal M}$ occurs,
 the intermediate step passing through a state within ${\cal M}$ speeds up the melting of the nonzero order parameter and, therefore, provides a first realization of PME, although typically 
 not very efficient. However, one can devise an alternative protocol where the DPT and the corresponding metastable region ${\cal M}$ instead \emph{secures} an efficient PME. To that end, consider ${\bf S}$ and ${\bf F}$ to be states belonging to different ordered phases. The direct step ${\bf S}\to {\bf F}$ must pass through a DPT and thus is slowed down by the existence of a metastable region ${\cal M}$.  One can now use a two-step protocol to circumvent the region ${\cal M}$ by first letting the system evolve toward an auxiliary state {\bf A} in the disordered
phase, and then from a state {\bf I} (in the disordered phase) to the target state {\bf F}.
Both these steps proceed without encountering a DPT and hence the two-step protocol is much faster than the direct protocol. We thus arrive at an efficient PME by making a detour around the DPT region. While  DPTs are extremely useful for engineering efficient PME protocols, the associated long time region ${\cal M}$ renders standard single-step Mpemba protocols useless. Indeed, if a quantum Mpemba effect exists between states belonging to different ordered phases connected by a DPT, the corresponding time saving will effectively be nullified by the long time needed for traversing ${\cal M}$, \new{see App.~\ref{appD} for a detailed discussion.}

\new{The remainder of this paper is structured as follows. In Sec.~\ref{sec2}, we introduce the lattice model studied in this work as well as our self-consistent Lindblad approach.  For details, see App.~\ref{appA} and App.~\ref{appB}.  In Sec.~\ref{sec3}, we describe the phase diagram and possible DPTs in quench protocols, see also App.~\ref{appC} for further details.  The connection between DPTs and the PME is presented in Sec.~\ref{sec4}, and in Sec.~\ref{sec5}, we offer some concluding remarks.
Technical details and additional results can be found in the Appendix.
}
 
\section{Model and Lindblad approach}\label{sec2}

For concreteness, we study the lattice version of a  1D interacting electronic system describing the Peierls transition in conducting polymers \cite{Brazovskii1981,Brazovskii1984,Saxena1987}, whose rich phase diagram exhibits ordered phases characterized by order parameters with different real-space symmetries. 
The Hamiltonian for a system with $L$ sites (periodic boundary conditions) is  
\begin{equation}
H = \sum_{j=1}^L \left [ - ( J + \Delta_j )   \left ( c_{j}^\dagger c^{}_{j+1} + 
{\rm h.c.} \right) - \mu  c_{j}^\dagger c_{j}^{}
+ \frac{\Delta_j^2}{2 g^2} \right],
\label{s.2.1}
\end{equation}
with spinless fermion annihilation operators $c_j$ and the real-valued lattice displacement field \new{$\Delta_j$  \cite{Brazovskii1981,Brazovskii1984,Saxena1987}}. Here $J$ denotes the bare hopping strength, 
$g$ the coupling between fermions and displacement fields, and $\mu$ the chemical potential. In the continuum limit, Eq.~\eqref{s.2.1} is equivalent to the 1D GN model widely used in high-energy physics as paradigm for asymptotic freedom \cite{Gross1974,Affleck1982,Wolff1985,Pausch1991,Schnetz2004,Thies2006},
\new{see App.~\ref{appA}.}

For computing the dynamics of the open system, where fermions also couple to an external reservoir, we resort to the Lindblad approach \cite{Lindblad1976,Breuer2007} which efficiently describes the time evolution toward steady states \cite{Guimaraes2016,Nava2021,Nava2023,Cinnirella2024,Cinnirella2025}. 
To obtain the order parameters characterizing the various phases of the model in Eq.~\eqref{s.2.1}, 
we improve and extend the time-dependent SCMF method 
introduced in Ref.~\cite{Peronaci2015} for closed superconducting systems and generalized in Refs.~\cite{Nava2019,Nava2023_s,Nava2024} to open systems. 
Within the SCMF approximation, $\Delta_j(t)$ is determined by time-dependent
self-consistency equations,
\begin{equation}
\Delta_j (t) =  g^2 \left [ \theta_{j,j+1} (t)  + \theta_{j+1,j} (t) \right] ,
\,\, \theta_{j,j'}(t) = {\rm Tr}[\rho (t)  c_j^\dagger  c^{}_{j'}].
\label{s.2.2}
\end{equation}
\noindent
The time-dependent system density matrix $\rho (t)$ is obtained by solving the Lindblad equation \cite{Lindblad1976,Breuer2007} ($\hbar=1$), 
\begin{eqnarray}\nonumber
 \frac{d\rho(t)}{dt} &=& - i [ H (t) ,\rho (t) ]  + \gamma \sum_{\epsilon_t} 
\biggl ( [1-f(\epsilon_t)]\, {\cal D}\left[\Gamma_{\epsilon_t}^{}\right]\rho(t) \\ 
& +& f(\epsilon_t) \, {\cal D}\left[\Gamma^\dagger_{\epsilon_t}\right]\rho(t)\biggr) \label{tlme.2},
\end{eqnarray}
with the dissipator ${\cal D}[\Gamma]\rho\equiv \Gamma\rho \Gamma^\dagger-\frac12 \{ \Gamma^\dagger\Gamma,\rho\}$ and the anticommutator $\{\cdot,\cdot\}$, where 
 $H(t)$ is given by Eq.~(\ref{s.2.1}) with $\Delta_j (t)$ in Eq.~(\ref{s.2.2}), and $\epsilon_t$ denotes the instantaneous eigenvalues of $H(t)$. The 
 jump operators  $\Gamma_{\epsilon_t}$ and $\Gamma_{\epsilon_t}^\dagger$ correspond to the addition or removal of a fermion in the associated single-particle eigenmode from or into a fermionic environment, respectively, \new{see Eq.~\eqref{Gammadef} in App.~\ref{appB}}. Physically, this model for the environment represents, for instance, 
the effect of quasiparticle tunneling between the system and a tunnel-coupled substrate (e.g., a metallic gate) in the Markovian limit \cite{Nava2024}. 
The associated jump rates are encoded by an overall rate constant $\gamma$ and by Fermi function factors with $f (\epsilon) = 1/[1+e^{\epsilon/k_B T}]$. Following standard arguments \cite{Breuer2007}, Eq.~\eqref{tlme.2} applies at finite temperature $T$ and weak coupling $\gamma$. 

\new{Throughout, we use $J=1$ as energy scale, setting $\gamma=0.01$ and $k_B T=0.05$ to ensure validity of the Lindblad approach. For $\gamma>0$,  numerical simulations based on Eq.~\eqref{tlme.2} are stable and converge to the steady state on a time scale $\propto \gamma^{-1}$.  The chosen temperature 
puts us in the low-temperature regime where quantum effects are pronounced and sharp DPTs exist.  In particular, the phase diagram found for $k_B T=0.05 J$ from our approach reproduces earlier zero-temperature theoretical predictions, see Sec.~\ref{sec3} below.
We emphasize that the Lindblad equation \eqref{tlme.2} is nonlinear and effectively time-dependent because of the self-consistency condition \eqref{s.2.2}.  As a consequence, it is difficult to achieve analytical progress for the quench protocols under study here.  }

\section{Phase diagram and DPTs}\label{sec3}

Before turning to time-dependent protocols, let us address the phase diagram of this model. After initializing $\rho(t=0)$ in a random state, at $t=0^+$, the parameters are quenched to $(\mu,g)$ and the rate $\gamma$ is switched on.  The asymptotic long-time state $\rho(t\to\infty)$ obtained by solving  Eq.~\eqref{tlme.2} determines the equilibrium steady state, where $\{\Delta_j (t \to \infty)\}$ in 
Eq.~\eqref{s.2.2} yields the spatial order parameter profile. 
By collecting numerical results for $\rho(t\to \infty)$ with different $(\mu,g)$ at fixed $(\gamma,T)$, 
we map out the phase diagram in the $\mu$--$g$ plane.  Writing \cite{Brazovskii1981,Brazovskii1984,Saxena1987}
\begin{equation}\label{orderp}
    \Delta_j = \delta J + (-1)^j m_j,
\end{equation}
the uniform contribution $\delta J$ (which is perturbative in $g^2$)  provides 
a renormalization of $J$ which is kept implicit in what follows. 
Different phases are then distinguished by the order parameter profile $m_j$. 

\begin{figure}
\includegraphics[width=0.48\textwidth]{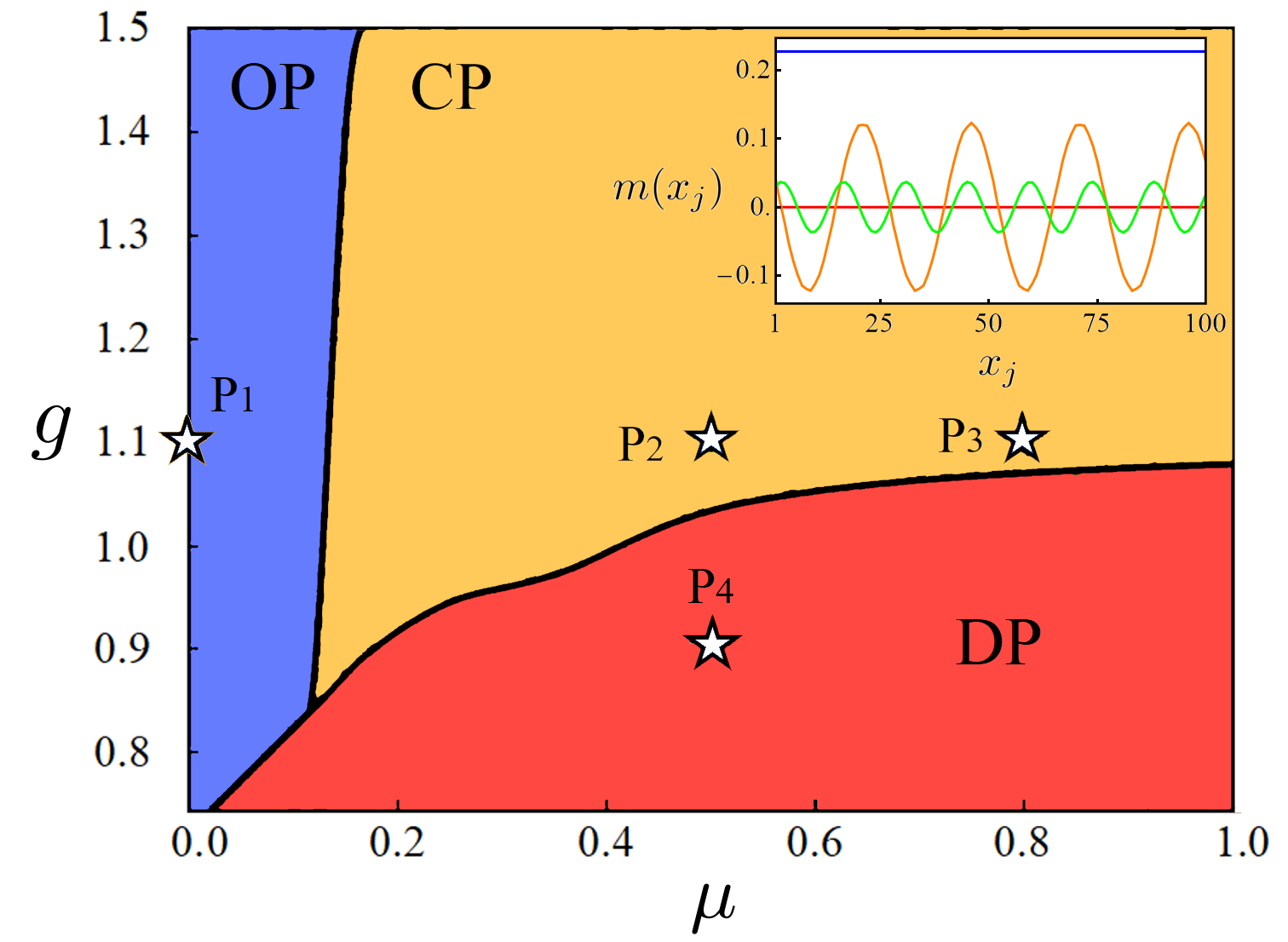}
\caption{Phase diagram of the model (\ref{s.2.1}) in the $\mu$--$g$ plane for $J=1$, $k_B T=0.05$, and $\gamma=0.01$.  The four points $P_i= (\mu_i, g_i)$ marked by stars correspond to $P_1=(0, 1.1)$,  $P_2=(0.5,1.1)$, $P_3=(0.8,1.1)$, and $P_4=(0.5,0.9)$, respectively.
 Results were obtained from the steady-state limit of Eq.~\eqref{tlme.2}.  
 The phases OP (blue), CP (orange), and DP (red) correspond to the ordered phase, crystal phase, and disordered phase, respectively; for details, see main text. 
Inset: Order parameter profile $m(x_j)$ at site $x_j=ja$ (with $a=1$), see Eq.~\eqref{orderp}, for the four points $P_i$ at system size $L=100$.
The blue curve corresponds to $P_1$, the red curve to $P_4$, and the orange and green curves to $P_2$ and $P_3$, respectively.  }  \label{fig1}
\end{figure}

\new{The equilibrium steady state value of the order parameter \eqref{orderp} can be determined with no need to resolve the complicated quench dynamics in Eq.~\eqref{tlme.2}. In the main panel of Fig.~\ref{fig1}, we show the corresponding phase diagram derived for system size $L=2000$. We have checked that the diagram is not affected when further increasing $L$, which allows us to extrapolate our results for the phase diagram to the thermodynamic limit.} Specifically, by analyzing the order parameter profile $m_j$ as shown in the inset, we identify three different phases, namely (i) an ordered  phase (OP) at small values of $\mu$, with finite and uniform 
 $m_j=m\ne 0$, (ii) a disordered phase (DP) at small $g$, with vanishing order parameter $m_j=0$, 
 and (iii) a crystal phase (CP) with a periodic modulation of $m_j$.  
 The phase diagram in Fig.~\ref{fig1} \new{computed for $k_BT =0.05J$} is consistent with the \new{zero-temperature} phase diagram derived in Refs.~\cite{Brazovskii1984,Wolff1985,Pausch1991,Thies2006}, but here is obtained by the simpler route of numerically solving the Lindblad equation \eqref{tlme.2} with the time-dependent SCMF approximation. In the inset of Fig.~\ref{fig1}, we show the steady-state profiles  $m_j$ for the four points marked by a star in the main panel. 
The blue curve (OP) shows a constant profile, $m_j=m\ne 0$, the orange and  green curves (CP) show a periodic modulation with momenta $Q=2 \pi \nu/L$ for $\nu=4$ and $\nu=7$, respectively, while the red curve (DP) gives $m_j=0$.

\begin{figure}
\includegraphics[width=0.48\textwidth]{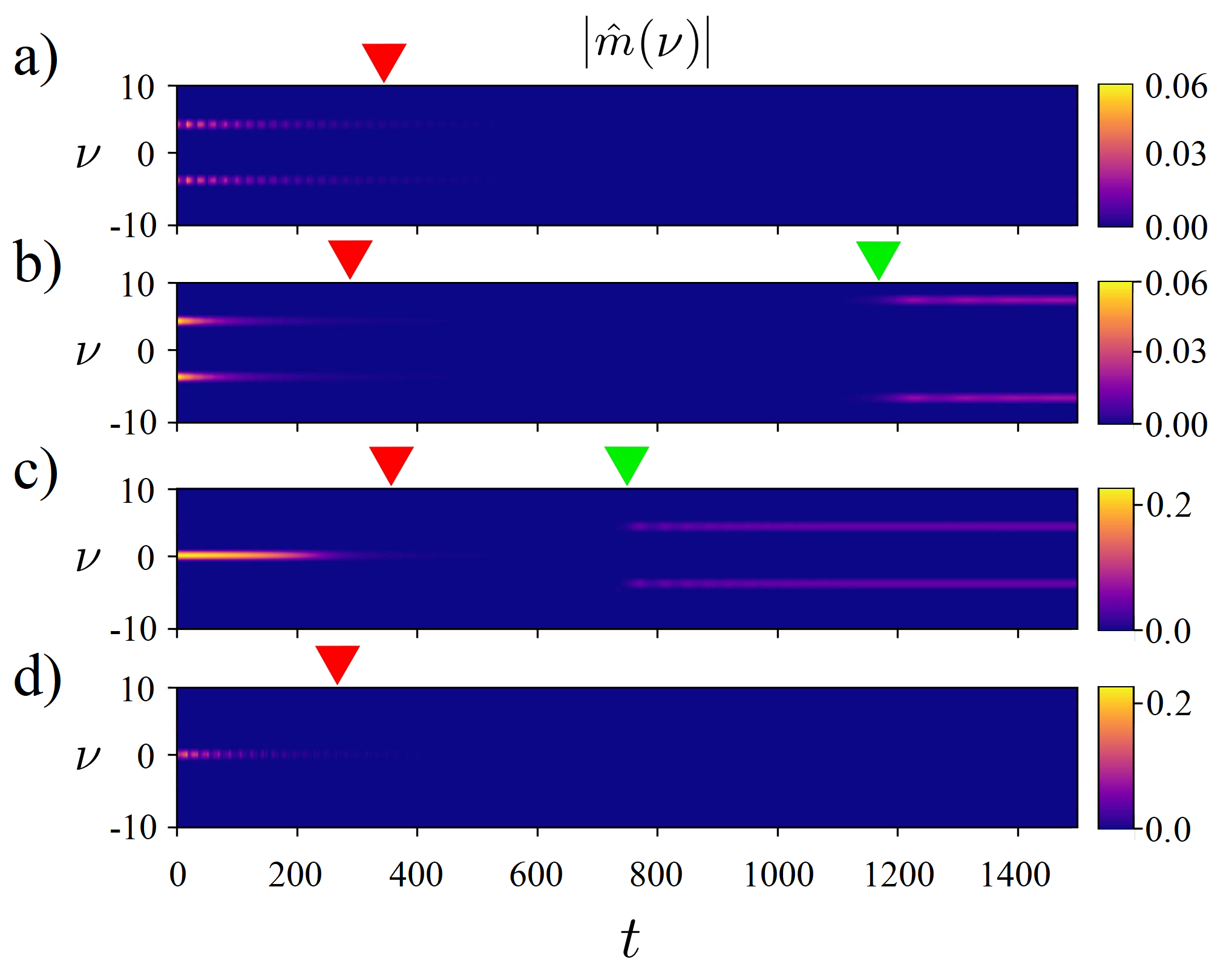}
\caption{Color-scale plot for the time evolution of the lowest 21 Fourier modes $\hat{m}(\nu,t)$ of the order parameter $m_j$ in Eq.~\eqref{orderp} under parameter quenches between different regions of the phase diagram in Fig.~\ref{fig1}. We use $L=100$ and $(\gamma,T)$ as in Fig.~\ref{fig1}.
Green arrows mark the critical time $t_*$ corresponding to DPTs. Red arrows mark the time scales for a relaxation crossover. Different panels correspond to (see main text for details): 
(a) Quench from CP to DP. (b) Quench between two states in the CP. (c) Quench from OP to CP. (d) Quench from OP to DP. }
\label{fig2}
\end{figure}

The phase diagram in Fig.~\ref{fig1}, containing ordered phases (OP, CP) with different order parameter symmetries, resembles the one discussed in Refs.~\cite{Nava2023_s,Nava2024} for  planar superconductors. From the results of Refs.~\cite{Nava2023_s,Nava2024}, we then infer that a quench between different ordered phases must trigger a DPT at some finite critical 
time $t_*$. To induce a DPT in our model, we adapt the protocol in Refs.~\cite{Nava2023_s,Nava2024}: At $t=0^+$, the parameters are quenched from their initial values $P_{\rm in}=(\mu_{\rm in}, g_{\rm in})$  to the final values
$P_{\rm eq}=(\mu_{\rm eq}, g_{\rm eq})$. By numerically solving the coupled Eqs.~(\ref{s.2.2}) and (\ref{tlme.2}), we then obtain the time-dependent order parameter. In Fig.~\ref{fig2},  we show the time evolution of the lowest 21 Fourier harmonics $\hat m(\nu)$ of $m_j(t)$ with momentum $Q = 2 \pi \nu/L$ and integer $|\nu|\le 10$ (which amply suffices to capture all observed spatial profiles of $m_j$) for four different quench protocols using pairs of the four points $\{P_i\}$ in Fig.~\ref{fig1}.
In particular, Fig.~\ref{fig2}(a) shows the time evolution  
from $P_2\to P_4$ (CP $\to$ DP). While initially all spectral weight in the CP is contained in the harmonics with $\nu = \pm 4$, along the time evolution to the DP, these weights smoothly fade away and we arrive at a conventional relaxation process (without DPT) toward $\hat{m}(\nu)=0$.
Figure~\ref{fig2}(b) corresponds to  $P_{\rm in}=P_2$ and  $P_{\rm eq}=P_3$, where both states are in the CP. In this case, a DPT is observed since the order parameters have different periodicity.  Clearly, there is an extended time region ${\cal M}$ before the DPT occurs, $150\alt t \alt t_*\approx 1200$, where the order parameter weights spread over all Fourier harmonics, each one being very small. 
Next, in Fig.~\ref{fig2}(c), we study a quench from $P_{\rm in}=P_1$ to $P_{\rm eq}=P_2$ (OP to CP), where we again encounter a DPT separating both phases. 
The region ${\cal M}$ now extends over the time span $200 \alt t \alt t_*\approx 800$.
Finally, in Fig.~\ref{fig2}(d), for $P_{\rm in}=P_1$ and $P_{\rm eq}=P_4$ (OP to DP), 
again a relaxation dynamics as in Fig.~\ref{fig2}(a) is observed. In App.~\ref{appC}, we complement those results by monitoring the time evolution of each harmonic and the discontinuities in the time-dependent fidelity 
\cite{Zhang2022,Zhang2023,Strachan2025,Parez2025}.

\section{PME protocols}\label{sec4}

To realize the PME, one needs to specify the states in parameter space, 
$\{ {\bf S},  {\bf F},  {\bf A},  {\bf I}\}$. In addition, one has to define a suitable distance measure between quantum states $\rho$ and $\rho'$ \cite{Lu2017,Nava2025}. For small systems, a rigorous and physically meaningful measure is given, e.g., by the trace distance ${\cal D}_{\rho,\rho'} = \frac{1}{2} {\rm Tr} |\rho-\rho'|$ \cite{Nava2024m,Zatsarynna2025}, \new{but other measures such as the Bures distance (based on the fidelity) or the quantum relative entropy can also be employed.} 
In our case,  ${\cal D}_{\rho,\rho'}$ is impractical since the size of the Hilbert space becomes exponentially large in $L$ and the time-dependent SCMF approach renders the dynamics intrinsically nonlinear.  \new{However, since the system under study exhibits phase transitions,  it actually suffices to monitor the order parameter dynamics in order to reliably detect quantum Mpemba effects \cite{Nava2019}.} For these reasons, we here quantify the state distance  in terms of the \new{normalized} \emph{order parameter distance},  
\begin{equation}
\label{order_parameter_distance}
M(t)=\frac{\sqrt{\sum_\nu \left [ \hat{m}(\nu,t)-\hat{m}_{\rm eq}(\nu) \right ]^2}}{\sqrt{\sum_\nu \left [ \hat{m}(\nu,0)-\hat{m}_{\rm eq}(\nu) \right ]^2}},
\end{equation}
\noindent
where $\hat{m}(\nu,t)$ is the time-dependent Fourier mode of $m_j(t)$ in Eq.~\eqref{orderp} at momentum $Q = 2 \pi \nu/L$ and $\hat{m}_{\rm eq}(\nu)$ the corresponding steady-state value. 
According to Eq.~\eqref{order_parameter_distance}, we have $M(t=0)=1$ and  $M(t \to \infty)=0$.

\begin{figure}
\includegraphics[width=0.48\textwidth]{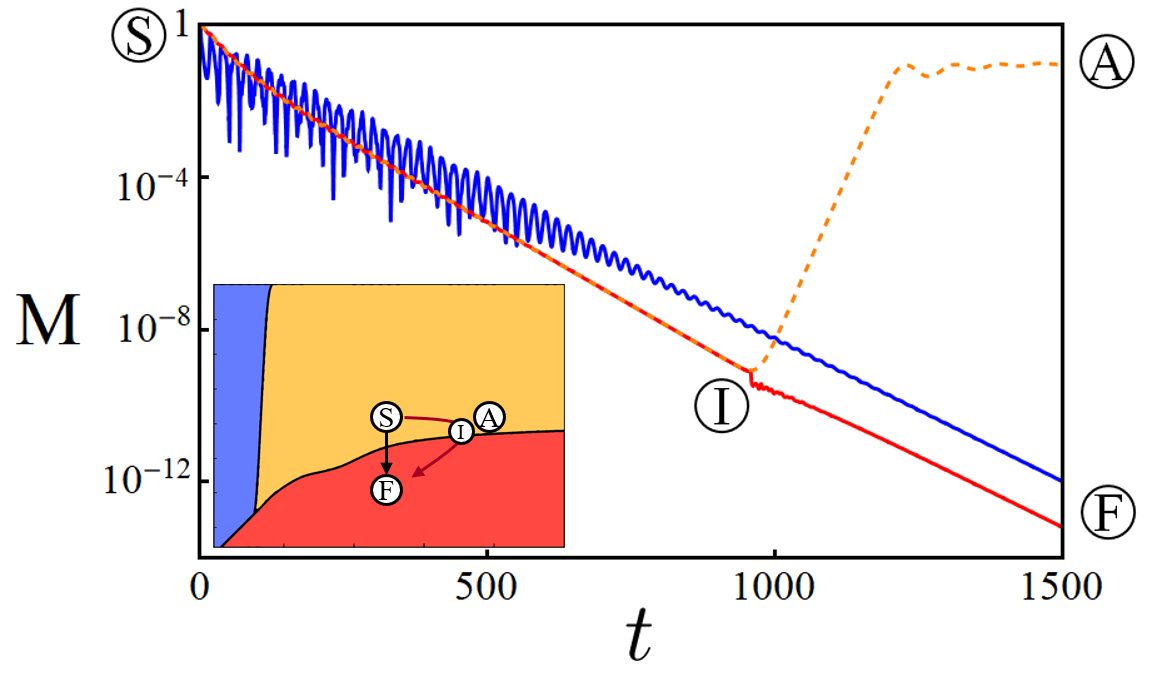}
\caption{PME for the GN model. Main panel: \new{Dimensionless (normalized) order} parameter distance $M(t)$ vs time $t$ \new{(in units of $1/J$)}, see Eq.~\eqref{order_parameter_distance}, computed from Eq.~\eqref{tlme.2} for two different protocols from ${\bf S}=P_2 \to {\bf F}=P_4$, see Fig.~\ref{fig1}.  Notice the semi-logarithmic scales. We use $L=100, \gamma=0.01,$ and $k_BT=0.05$.
The blue curve corresponds to the single-step direct quench ${\bf S}\to {\bf F}$.
The red curve corresponds to a two-step process, where the system first evolves along
${\bf S}\to {\bf A}=P_3$. At $t= 960$, the state {\bf I} is reached. Now a second quench 
takes the system from ${\bf I}\to {\bf F}$. The orange-dashed curve is for the single-step protocol ${\bf S}\to {\bf A}$. Inset: Location of the parameter states $\{{\bf S},{\bf F},{\bf A},{\bf I}\}$ in the phase diagram, see Fig.~\ref{fig1}.  The black curve indicates the direct step ${\bf S}\to {\bf F}$, the dark red curve the two-step protocol ${\bf S}\to {\bf I}\to {\bf F}$. Note that the states along these trajectories are actually nonequilibrium states.
}
\label{fig3}
\end{figure}

In order to select parameter configurations $\{ {\bf S},  {\bf F},  {\bf A},  {\bf I}\}$ for  PME protocols, we first recall the Fourier mode dynamics in Figs.~\ref{fig2}(b,c).  In both cases, there is a DPT and thus a long intermediate time region ${\cal M}$ exists during which the spectral weights $\hat{m}(\nu,t)$ slowly redistribute from just a few modes at short times to a broad continuum of harmonics.  
In Fig.~\ref{fig3}, we show $M(t)$ as obtained by solving Eq.~\eqref{tlme.2} after quenching the system parameters from ${\bf S} = P_2$ (CP) to ${\bf F}=P_4$ (DP), see Fig.~\ref{fig1}. 
The blue curve in Fig.~\ref{fig3} shows $M(t)$ for the direct step ${\bf S}\to {\bf F}$, while the red curve illustrates $M(t)$ for a two-step protocol using the 
auxiliary state ${\bf A}=P_3$ (CP), where the existence of a DPT implies an intermediate region ${\cal M}$. Here, ${\bf I}$ is chosen as the point of minimal distance from {\bf F} along the trajectory ${\bf S}\to  {\bf A}$, see Fig.~\ref{fig2}(b).  Even though the direct step here does not traverse a metastable region ${\cal M}$, the two-step process is still faster if the state 
${\bf I}$ is chosen wisely. 
In this example, by letting the system pass through ${\cal M}$ during the two-step process, one speeds up the relaxation, thus providing a first realization of PME.

\begin{figure}
\includegraphics[width=0.48\textwidth]{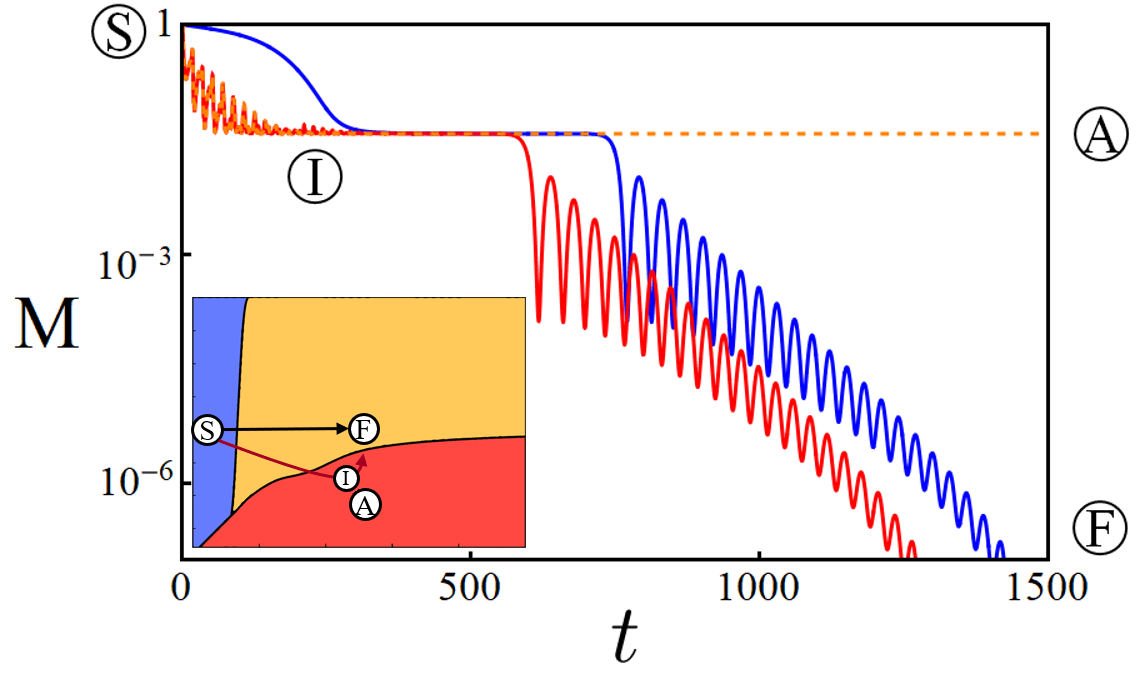}
\caption{Same as in Fig.~\ref{fig3} but for ${\bf S}=P_1, {\bf F}=P_2$ and ${\bf A}=P_4$. The intermediate point {\bf I} is reached at $t=200$ along the trajectory ${\bf S} \to {\bf A}$.
Main panel: The blue curve shows the dynamics under the single-step protocol ${\bf S}\to {\bf F}$, the orange-dashed curve is for a single-step evolution ${\bf S}\to {\bf A}$. The red curve refers to the two-step protocol ${\bf S}\to {\bf I}\to {\bf F}$.
Inset: Location of the parameter states $\{{\bf S},{\bf F},{\bf A},{\bf I}\}$ in the phase diagram, see Fig.~\ref{fig1}. The black curve indicates the direct step ${\bf S}\to {\bf F}$, the dark red curve the two-step protocol ${\bf S}\to {\bf I}\to {\bf F}$. Note that the states along these trajectories are actually nonequilibrium states. }
\label{fig4}
\end{figure}

\begin{figure}
\includegraphics[width=0.4\textwidth]{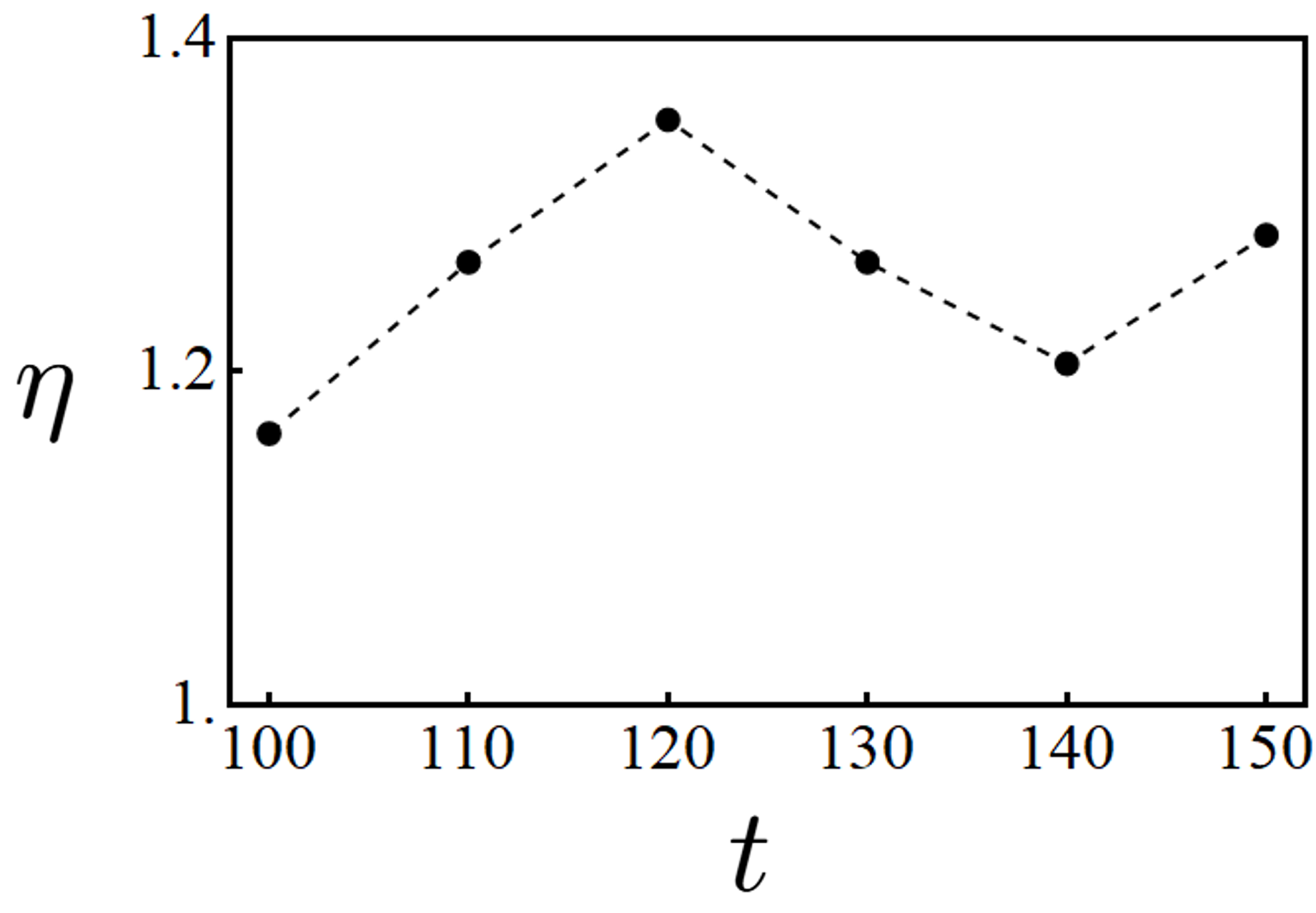}
\caption{ \new{PME speedup ratio $\eta$ vs system size $L$ for the protocol shown in the main panel of Fig.~\ref{fig3}. For details, see main text.} }
\label{fig5}
\end{figure}

A larger enhancement of the PME efficiency can be achieved by a different use of the DPTs as
shown in Fig.~\ref{fig4}, where we plot $M(t)$  after quenching
${\bf S} = P_1 \to {\bf F}=P_2$. We now select the state {\bf I} corresponding to the start of the plateau region along the trajectory ${\bf S}\to {\bf A}$. 
Again, the blue curve shows $M(t)$ for the direct protocol ${\bf S}\to {\bf F}$, 
while the red curve shows $M(t)$ along the two-step protocol employing ${\bf A}=P_4$. In this 
case, using the DP as the intermediate phase, both steps do not encounter a long metastable region ${\cal M}$ while the direct step has to traverse such a region, see Fig.~\ref{fig2}(c). 
By contrast, in the two-step protocol, the system evolves along the faster
OP~$\to$~DP crossover, see Fig.~\ref{fig2}(d), and along the (inverse)  CP~$\to$~DP crossover, see Fig.~\ref{fig2}(a).  As a result, we arrive at an efficient PME.

\new{One can quantify the speedup of the relaxation time in the PME by taking the ratio $\eta$ of the corresponding relaxation times, $\eta=\tau_{{\rm direct}}/\tau_{{\rm two-step}}$. 
The relaxation time $\tau$ is here defined by $|M(t>\tau)|<\epsilon_0$ with a small threshold value $\epsilon_0\ll 1$.  While the precise value of $\tau$ depends on $\epsilon_0$, the ratio $\eta$ is independent of this parameter.  
While based on our discussion in Sec.~\ref{sec1}, we do not expect universal scaling in the dependence of $\eta$ on system parameters, Fig.~\ref{fig5} illustrates the system size ($L$) dependence of $\eta$ obtained numerically for the protocol shown in the main panel of Fig.~\ref{fig3}.  We observe that $\eta$ is only weakly dependent on $L$ and always well above unity.   Let us also emphasize that here we have not optimized the two-step protocol but simply took the values from Fig.~\ref{fig3}. In fact, the connection between DPTs and the PME is robust and does not rely on fine-tuning of parameters. }

\section{Conclusions}\label{sec5}

In this \new{paper}, we have pointed out an intriguing interplay between Mpemba effects and DPTs, using the latter as an efficient way to gain control on the former. 
In doing so, we have developed a powerful approach to constructing the phase diagram of correlated electron models in terms of the Lindblad equation and a time-dependent SCMF approximation. Arguably, this approach has a wide range of applicability, e.g., to models
in condensed matter or high-energy physics as well as in quantum chemistry.  Deepening our analysis of the connection between Mpemba effects and DPTs, and extending our methods to other, possibly higher-dimensional, correlated fermion models are interesting topics for future research. 
\new{Our predictions may be experimentally tested in different platforms of current interest, including driven-dissipative superconducting qubits \cite{Rasmussen2021,Westhoff2025}, ultracold atoms \cite{Bloch2008}, or ion traps.  In particular, ion trap experiments have already been used for studying quantum Mpemba effects \cite{Joshi2024,Zhang2025}, and similar setups could allow to observe the predicted interplay between DPTs and Mpemba effects.
}
\section*{Data availability}

The data underlying the figures in this work can be found at Zenodo \cite{Zenodo}.

\begin{acknowledgments} 
We acknowledge funding by the Deutsche Forschungsgemeinschaft (DFG, German Research Foundation) under Projektnummer 277101999 - TRR 183 (projects B02 and C01), under Projektnummer EG 96/14-1, and under Germany's Excellence Strategy - Cluster of Excellence Matter and Light for Quantum Computing (ML4Q) EXC 2004/1 - 390534769.
\end{acknowledgments}


\appendix
\setcounter{figure}{0}
\renewcommand{\thefigure}{A\arabic{figure}}

\section*{APPENDIX}

\new{The Appendix is structured as follows.} 
In App.~\ref{appA}, we show how the low-energy continuum limit 
of the lattice model \eqref{s.2.1} leads to the 1D massless GN model.
In App.~\ref{appB}, we present our numerical implementation of the Lindblad approach, and in App.~\ref{appC}, we discuss the phase diagram of the lattice model. Finally, in App.~\ref{appD}, we study  implementations of the standard quantum Mpemba protocol in our system and compare them to the PME discussed in the main text.

\section{Continuum limit of the lattice model}
\label{appA}

We consider the lattice Hamiltonian $H$ in Eq.~\eqref{s.2.1} with the displacement fields $\Delta_j(t)$ as determined by Eq.~\eqref{s.2.2}.
To account for the staggered component $m_j$ of the displacement field, 
see Eq.~\eqref{orderp}, we divide the Brillouin zone into two parts by setting 
\begin{eqnarray}
c_j &=& \frac{1}{\sqrt{L}}\: \sum_{0\leq k \leq 2 \pi} c_k e^{ikj} \nonumber \\
&\approx&  \frac{1}{\sqrt{L}}\: \sum_{0\leq k \leq   \pi}\left( i^j c_k e^{ikj} + (-i)^j c_{k+\pi} e^{ikj} \right) \nonumber \\
 &\equiv& \frac{1}{\sqrt{2}}\left( i^j c_{1} (x_j ) + (-1)^j c_2 (x_j )\right), \label{sup.1.3}
 \end{eqnarray}
 with smooth operator functions $c_{1,2} (x)$. We then have
 \begin{eqnarray}
 &-&J   \sum_{j=1}^L \: \{ c_{j}^\dagger c_{j+1 } + 
c_{j+1}^\dagger c_{j} ]  \} -  \mu \sum_{j=1}^L \:  c_{j}^\dagger c_{j} 
 \nonumber \\
 &\approx& -iJ   \sum_{j=1}^L \left( c_{1}^\dagger (x_j) c_{1} (x_{j+1})  - c_{2}^\dagger (x_j) c_{2} (x_{j+1}) -
 {\rm h.c.}\right) \nonumber \\
  &-& \mu  \sum_{j=1}^L \left( c_{1}^\dagger (x_j) c_{1} (x_j )  + c_{2}^\dagger (x_j) c_{2} (x_j )  \right). 
 \label{sup.1.4}
 \end{eqnarray}
Similarly, we get 
 \begin{eqnarray}
 &-&  \sum_{j=1}^L  (-1)^j m(x_j) [ c_{j}^\dagger c_{j+1 } + 
c_{j+1}^\dagger c_{j} ]   \label{sup.1.5}\\
&\approx& - i \sum_{j}  m(x_j) [c_1^\dagger (x_j) c_2 (x_{j+1} )  - c_2^\dagger (x_j) c_1 (x_{j+1})] + {\rm h.c.} 
\nonumber 
\end{eqnarray}
Expanding the terms on the r.h.s.~of Eqs.~\eqref{sup.1.4} and \eqref{sup.1.5}, retaining only leading contributions in the lattice constant $a$ (where eventually $a=1$), and trading sums for integrals, we obtain 
\begin{eqnarray}
H &\to&  \int_0^L  d x \, \bigl\{ c_1^\dagger (x) [-iv \partial_x - \mu ]c_1 (x) \nonumber \\
&+& c_2^\dagger (x) [ iv\partial_x - \mu ]c_2 (x) + \frac{1}{2 g^2} [2m(x)]^2
\nonumber \\
&-& 2 i m (x ) [ c_1^\dagger (x) c_2 (x) - c_2^\dagger (x) c_1 ( x ) ] \bigr\},  
\label{sup.1.6}
\end{eqnarray}
with $v=2aJ$.
Equation~\eqref{sup.1.6} corresponds to the Hamiltonian  
used in Refs.~\cite{Brazovskii1981,Brazovskii1984,Saxena1987} to study the Peierls effect in conducting polymers, with the lattice potential $\Delta_e=0$ generated by the rigid polymer skeleton, 
the fields $\psi_\pm (x)= e^{\pm \frac{i \pi}{4} } c_{1,2} (x)$,
and the staggered potential $\Delta_i (x)= 2 m(x)$. In the continuum limit, our model then precisely coincides with the continuum model in Eq.~(1) of Ref.~\cite{Brazovskii1984}.

Equation~\eqref{sup.1.6} also corresponds to the 1D massless GN Hamiltonian $H_{\rm GN}$ \cite{Gross1974,Pausch1991} at finite chemical potential $\mu$. Indeed, with the bispinor $\psi (x) = \left(\begin{array}{c}
 \psi_+ (x) \\ \psi_- (x) \end{array} \right)$,
 we obtain from Eq.~\eqref{sup.1.6} the GN model, 
  \begin{eqnarray}
 && H \to  H_{\rm GN} = \int  d x   \biggl( \psi^\dagger (x) [ - \mu \sigma^0 - i v \sigma^z \partial_x ] \psi (x) 
 \nonumber \\
 &+& \Delta (x) \, \psi^\dagger (x) \sigma^y \psi (x)  + \frac{1}{2g^2} \Delta^2 (x) \biggr) ,
 \label{sup.1.10}
 \end{eqnarray}
 with $\Delta(x)=2m(x)$, the Pauli matrices $\sigma^{x,y,z}$ and the identity $\sigma^0$. The standard representation of $H_{\rm GN}$ used in high-energy physics \cite{Gross1974,Pausch1991} follows from Eq.~\eqref{sup.1.10} after a unitary transformation.
 We note that for a multi-flavor generalization, one adds an extra flavor index $\alpha$ such that $c_j \to c_{j,\alpha}$, where $H_{\rm GN}$ then contains a sum over $\alpha$.  
   
 \section{Lindblad approach}
 \label{appB}

 We here address the numerical solution of the Lindblad equation. We compute $\rho (t)$  from Eq.~\eqref{tlme.2}, where the time dependence of 
$H(t)$ stems from $\Delta_j(t)$ via the self-consistency equation \eqref{s.2.2}, see also Refs.~\cite{Nava2023_s,Nava2024}.  After discretizing time on a sufficiently fine grid, 
at a given time, we diagonalize $H(t)$, resulting in
the eigenvalues $\epsilon_t$ and the \new{associated quasiparticle eigenmodes,
\begin{equation}\label{Gammadef}
\Gamma_{\epsilon_t}= \sum_{j=1}^L u_{\epsilon_t,j}^* c_j,
\end{equation} 
with complex-valued coefficients $u_{\epsilon_t,j}$ forming the respective eigenvector.  The jump operators in the Lindblad equation \eqref{tlme.2} were chosen as $\Gamma^{}_{\epsilon_t}$ and $\Gamma^\dagger_{\epsilon_t}$, corresponding to the exchange of a single quasiparticle with the weakly coupled reservoir.
By inverting Eq.~\eqref{Gammadef}} and using the correlation matrix 
$\theta_{j,j'}(t)$ in Eq.~\eqref{s.2.2}, the Lindblad equation together with the self-consistency condition \eqref{s.2.2} describes an intrinsically nonlinear dynamics. 

However, if one is interested in the order parameter $m_j(t)$ only, see Eq.~\eqref{orderp}, a more direct way is to solve a closed set of differential equations for $\theta_{j,j'} (t)$. Omitting the time argument $t$ in both $\sigma_j(t)$ and $\theta_{j,j'}(t)$ for notational simplicity, we obtain
\begin{widetext}
\begin{eqnarray}\label{s.2.11}
&&\frac{d \theta_{j,j'}}{d t }= - i ( J + \Delta_{j -1} ) \theta_{j-1,j'} + 
i ( J + \Delta_{j'})  \theta_{j, j'+1 } 
-i (J+\Delta_j)  \theta_{j+1,j'} + i ( J + \Delta_{j'-1} )\theta_{j,j'-1} + \\\nonumber 
&& +\frac{\gamma}{2} \sum_{\epsilon_t} \sum_{r=1}^L \left\{- [1-f (\epsilon_t )] \: 
 \left( u_{\epsilon_t,r} u_{\epsilon_t,j}^* \theta_{r,j'}  
+ u_{\epsilon_t,j'} u_{\epsilon_t,r}^* \theta_{j,r} \right) + f (\epsilon_t ) \: 
\left( u_{\epsilon_t,r}^* u_{\epsilon_t,j'} [\delta_{r,j} - \theta_{j,r}  ] 
+  u_{\epsilon_t,j}^* u_{\epsilon_t,r} [\delta_{j',r}  - \theta_{r,j'} ] \right) \right\}. 
\end{eqnarray}
\end{widetext}
From the steady-state values $\Delta_j (t \to \infty)$ and Eq.~\eqref{orderp},
we then obtain the corresponding steady-state order parameter profile $m_j$.

In Refs.~\cite{Nava2023_s,Nava2024}, it has been shown that this approach works
for planar superconductors with $(s,d,id)$-wave order parameters, 
including combinations of two of those pairing symmetries. 
Here we have generalized this scheme by lifting all constraints on the dependence of $m_j$ on the site index $j$.
 For a numerical integration of Eq.~(\ref{s.2.11}) together with Eq.~\eqref{s.2.2}, we choose random initial conditions for $\theta_{j,j'} (t=0)$.  The fact that we find that the steady state is independent of the initial conditions lends further support to our approach. 

Let us then summarize general constraints on the correlation matrix $\theta_{j,j'}(t)$. First, 
all eigenvalues must be non-negative since they correspond to occupation numbers 
of the eigenmodes of $H(t)$. Second, its diagonal elements must satisfy  $\theta_{j,j} (t) \leq 1$ at all times. 
Similarly, we have $\sum_{j=1}^L \theta_{j,j} (t) = \bar{N} (t)$, with the average total particle number $\bar{N}(t)$ at time $t$. Third, in order to avoid that the time-evolving system gets trapped in a restricted subset of all possible configurations, we first define an $L \times L$ diagonal matrix ${\bf D}$ such that the initial filling is set at $\bar{N}/L = \frac{1}{2}$, i.e., 
$\frac{1}{L} \sum_{j=1}^L D_{j,j} =\frac{1}{2}$.
Next, we perform a basis  change by applying a unitary transformation ${\bf U}$, i.e.,
${\bf D} \to {\bf C} = {\bf U}^\dagger {\bf D} {\bf U}$.  
The correlation matrix $\theta_{j,j'}(t)$ has no translational symmetries, neither in the occupation probabilities (diagonal) nor in the order parameter (second diagonal), while ${\bf U}$ defines the basis where ${\bf D}$ is diagonal. In order to apply a small perturbation, we write ${\bf U} = {\bf 1} + i \epsilon {\bf A}$, with   a random Hermitian matrix ${\bf A}$ and 
$0< \epsilon \ll 1$.  Putting $\theta(0)={\bf C}$, we then let the correlation matrix evolve according to Eq.~(\ref{s.2.11}). Since at this stage, we are not interested in the time evolution but only in the steady state, the precise value of the system-environment coupling $\gamma$ is not relevant 
provided (as we have carefully checked) that $\gamma$ is both strong enough to equilibrate the system on a reasonably short time scale $\tau_\gamma \propto \gamma^{-1}$, yet not too strong to invalidate the derivation of the Lindblad equation \cite{Breuer2007} and/or to change the steady state.  In fact, for sufficiently large $\gamma$, the steady state explicitly depends on $\gamma$.  Specifically, we set $\gamma = 0.01J$ and let the
 system evolve until it reaches the steady state. In doing so, for fixed $J=1$ and $k_B T$, we sequentially scan through the parameters $\mu$ and $g$, spanning a grid in the $\mu$--$g$ plane. The grid is chosen to be sufficiently dense to yield accurate phase boundaries, see Fig.~\ref{fig1}. We monitor the order parameter $m_j$ both in real space and in momentum space, where the spectral weight $\hat{m}(\nu)$ is evaluated for all  modulation momenta $Q=2\pi \nu /L$ with integer $\nu$. 

\section{Phase diagram and DPTs}
\label{appC}
 
By employing the time-dependent SCMF approximation in the Lindblad approach and monitoring how
the steady-state order parameter profile $m_j$ depends on $\mu$ and $g$, we obtain the phase 
diagram in Fig.~\ref{fig1}.  In the \new{low-temperature limit} and at small $\mu$, the system is asymptotically free and in an ordered phase (OP) with uniform $m (x) = m_*\ne 0$,
with $m_*  \approx \pi J e^{-\pi J/(2 g^2)}$. On increasing $\mu$ while keeping $g$ fixed
with $g < g_{\rm cr} \approx 0.85 J$,  the system goes through a first-order 
phase transition at $\mu \sim m_*$ toward the disordered phase (DP) with $m(x) = 0$.
In the DP, the dynamics is as for a noninteracting electron chain with slightly renormalized parameters \cite{Wolff1985,Pausch1991}. 
For $g>g_{\rm cr}$,  the OP instead evolves into a modulated crystal phase (CP)
 \cite{Brazovskii1981,Brazovskii1984,Pausch1991,Thies2006}, where 
 $m (x)\ne 0$ is modulated in real space.

\begin{figure}
\includegraphics[width=0.36\textwidth]{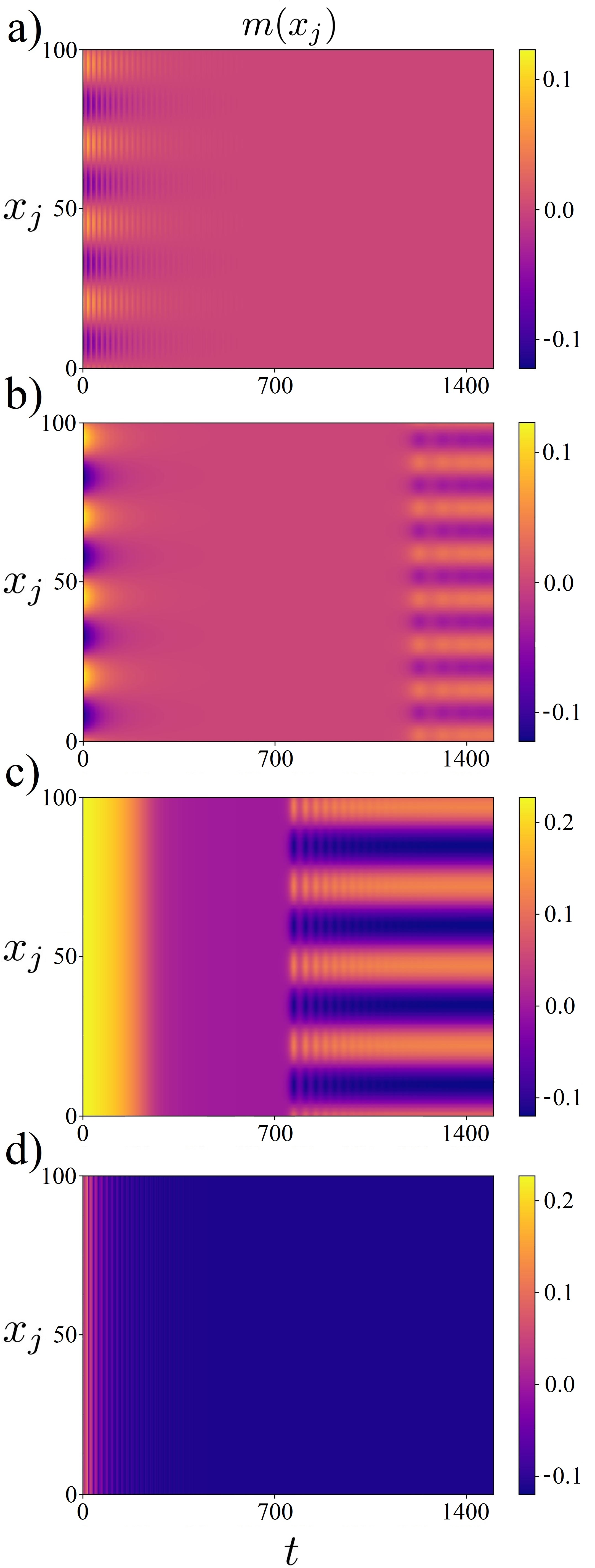}
\caption{Color-scale plots of $m(x_j,t)$ with $x_j=ja$ ($a=1$) for four different quench protocols $(\mu_{\rm in},g_{\rm in})\to (\mu_{\rm eq},g_{\rm eq})$, using $J=1$, $L=100$, $\gamma=0.01$ and $k_B T=0.05$.  Different panels correspond to (see text for details): (a) Quench from  CP to DP, (b) quench between two states  within the CP, (c) quench from OP to CP,  (d) quench from OP to DP.}
\label{figA1}
\end{figure}

\begin{figure}
\includegraphics[width=0.36\textwidth]{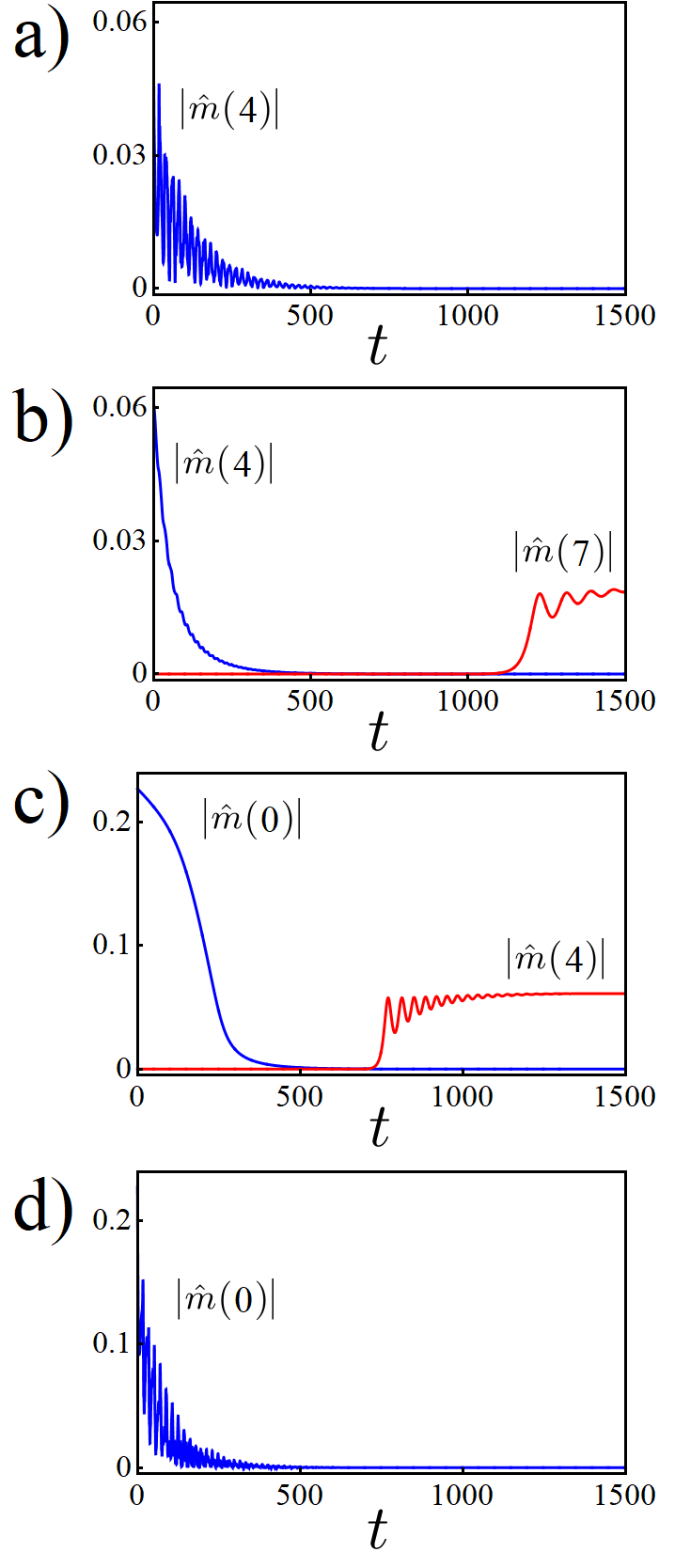}
\caption{Time evolution of the dominant harmonics $\hat{m}(\nu,t)$, see Eq.~\eqref{harmonics}, for initial and final states (blue and red curves, respectively) of the four quench protocols in Fig.~\ref{figA1}.  (a) Quench from CP to DP, (b) quench between two states within the CP,
(c) quench from OP to CP, (d) quench from OP to DP.}
\label{figA2}
\end{figure}
 
Following Refs.~\cite{Nava2023_s,Nava2024}, we next discuss DPTs between the various phases. 
To that end, we consider the numerical results in Figs.~\ref{figA1} and  \ref{figA2}.
In Fig.~\ref{figA2}, we show $m(x,t)$ for the same quench protocols $(\mu_{\rm in},g_{\rm in})\to (\mu_{\rm eq},g_{\rm eq})$ discussed 
in the main text. In particular, in the four panels of Figs.~\ref{figA1} and \ref{figA2}, we study the cases (a)  $P_2 \to P_4$ (CP to DP),  (b) $P_2\to P_3$ (both states in CP), (c) $P_1\to P_2$ (OP to CP), and (d) $P_1\to P_4$ (OP to DP), where the points $P_{1,2,3,4}$ defining $(\mu,g)$ correspond to the stars in Fig.~\ref{fig1}. 
In Fig.~\ref{figA2}, we instead plot the corresponding time evolution of the dominant Fourier harmonics, 
\begin{equation}\label{harmonics}
    \hat{m} (\nu,t) = \frac{1}{L}\sum_{j=1}^L e^{-\frac{2 \pi i \nu j}{L} } m_j(t),\quad 
\nu = -\frac{L}{2} , \ldots , \frac{L}{2}-1,
\end{equation}
for the initial and final states in (a)--(d).

In Fig.~\ref{figA1}(a),  we observe a continuous time decay of 
$m(x,t)$ from a modulated spatial dependence at short times (CP) to a uniformly vanishing value (DP), without evidence for a DPT. This is consistent with the results in Fig.~\ref{figA2}(a). We thus encounter a standard relaxation crossover, where the dominant Fourier harmonics at $\nu=\pm 4$ (in the CP) decay to zero.  A similar conclusion applies to 
the case shown in Figs.~\ref{figA1}(d) and \ref{figA2}(d), 
where we address the relaxation crossover from the OP to the DP. 
However, a DPT appears in Figs.~\ref{figA1}(b) and \ref{figA2}(b), where we study a quench between two points within the CP. In Fig.~\ref{figA1}(b), we identify a clear spatial modulation pattern at short times, $t\alt 50$, and again at long times, $t> t_*\approx 1200$. The shaded intermediate metastable region ${\cal M}$ covers
a wide time window in between, where no modulation with well-defined sharp periodicity exists.
Instead, we find a rather broad continuum of Fourier modes. 
Correspondingly, in Fig.~\ref{figA2}(b), after the quench, 
the $\nu=\pm 4$ harmonics continuously decay to zero. In the time region ${\cal M}$,  both dominant harmonics characterizing the initial $[\hat{m}(\pm 4)]$ and the
final [$\hat{m} (\pm 7)$] state become extremely small, and the spectral weight is
uniformly distributed across the whole momentum range, i.e., essentially all $\nu$ in Eq.~\eqref{harmonics} become important. As a consequence, no well-defined 
real-space modulation is visible in ${\cal M}$.  Finally,
for $t> t_*$, i.e., after the DPT, a finite amplitude $\hat{m} (\pm 7)$  corresponding  to the final CP state develops.
Similarly, a wide intermediate region ${\cal M}$ also appears for a quench from
OP~$\to$~CP as illustrated in Figs.~\ref{figA1}(c) and \ref{figA2}(c).

 \begin{figure}
\includegraphics[width=0.46\textwidth]{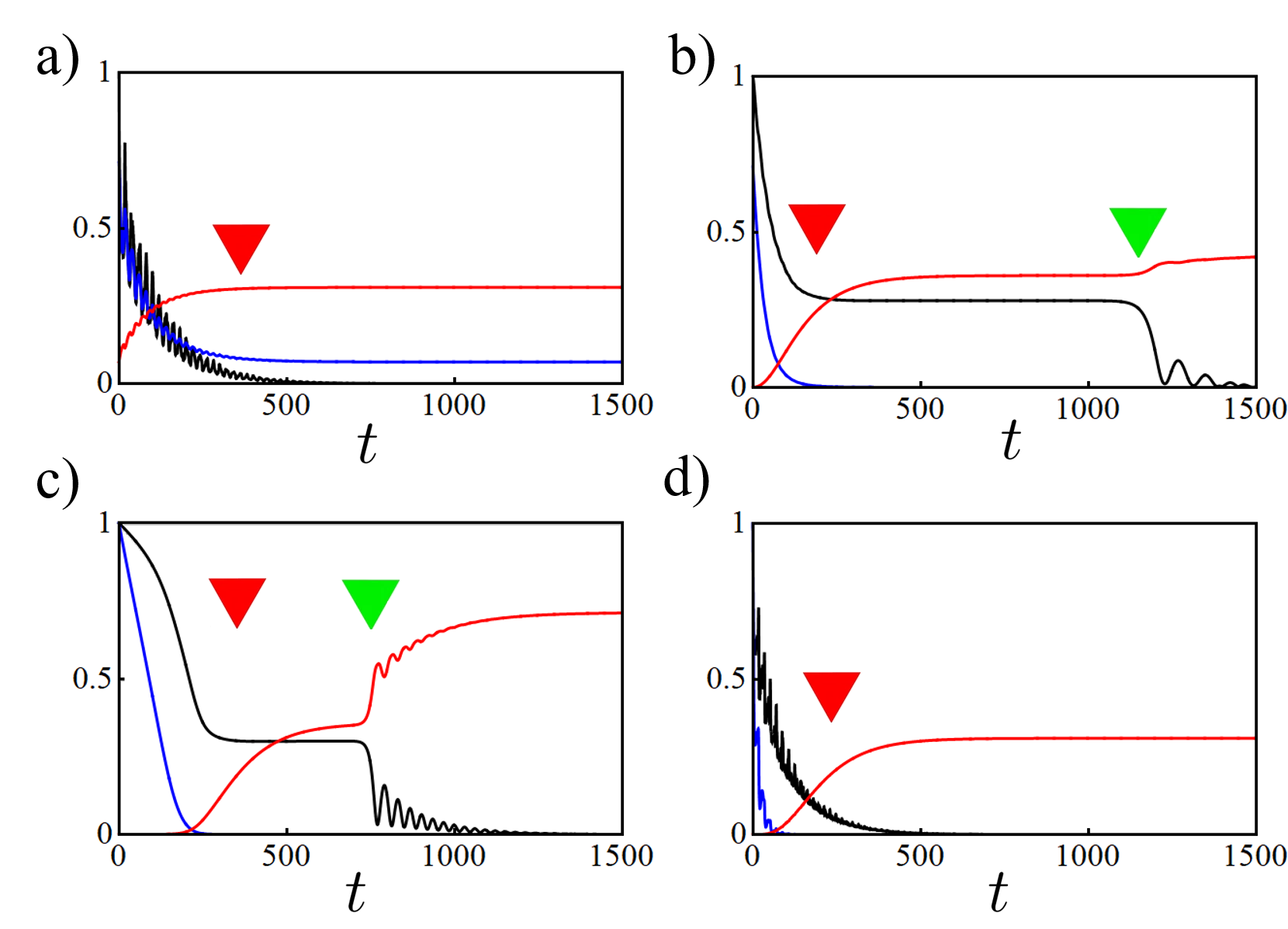}
\caption{Forward (backward) fidelity $F_{\rm fw}$ ($F_{\rm bw}$) vs time $t$, see Eq.~\eqref{fidely}, shown as blue (red) curves for four different 
quench protocols $(\mu_{\rm in},g_{\rm in})\to (\mu_{\rm eq},g_{\rm eq})$, using
again $J=1$, $L=100$, $\gamma=0.01$, and $k_B T = 0.05$.  Black curves refer to the time-dependent trace distance  ${\cal D}_T [\theta (t) ,\theta_{\rm eq}]$ in Eq.~(\ref{s.4.3}), normalized to its value at $t=0$. Red and green triangles mark relaxation crossovers and DPTs, respectively, as in Fig.~\ref{fig2}. 
Again the four panels correspond to (see text): (a) Quench from CP to DP, (b) quench between two states within the CP, (c) quench from OP to CP, (d) quench from OP to  DP.}
\label{figA3}
\end{figure}

To double check our conclusions, we also computed the time-dependent 
forward and backward fidelities, defined as
\begin{equation}
F_{\rm fw}[\rho(t),\rho_{\rm in}]={\rm Tr}[\rho(t) \rho_{\rm in}], \quad
F_{\rm bw}[\rho(t),\rho_{\rm eq}]={\rm Tr}[\rho(t) \rho_{\rm eq}],
\label{fidely}
\end{equation}
with $\rho_{\rm in} = \rho(t=0)$ and $\rho_{\rm eq} = \rho (t \to \infty)$. 
Both $F_{\rm fw}(t)$ and $F_{\rm bw}(t)$ take values in the interval $[0,1]$.
Roughly speaking, $F_{\rm fw}(t)$ [$F_{\rm bw}(t)$] indicate the ``distance'' between the states $\rho (t)$ and $\rho_{\rm in}$ [$\rho_{\rm eq}$].
Importantly, they can be expressed in terms of the correlation matrix $\theta_{j,j'}(t)$ in Eq.~\eqref{s.2.2}, see Refs.~\cite{Zhang2022,Zhang2023},
\begin{eqnarray}
F_{\rm fw}(t)&=& {\rm det}\left[ \mathds{1} - \theta (t) - \theta_{\rm in} + 
2 \theta (t) \theta_{\rm in} \right],
\nonumber \\
F_{\rm bw}(t)&=& {\rm det} \left[ \mathds{1}  - \theta (t) - \theta_{\rm eq} + 2 \theta (t) \theta_{\rm eq} \right], 
\label{s.4.1}
\end{eqnarray} 
with $\theta_{\rm in} = \theta(t=0)$ and $\theta_{\rm eq} = \theta(t \to \infty)$.
In Fig.~\ref{figA3}, we show the time dependence of $F_{\rm fw, bw}$ for the four quench protocols displayed in Figs.~\ref{figA1} and \ref{figA2}. 
With increasing time, $F_{\rm fw}(t)$ decays to 
zero in a similar way as the respective initially dominant harmonic mode shown in Fig.~\ref{figA2}. While $F_{\rm bw}(t)$ increases without sharp features in 
Figs.~\ref{figA3}(a,d), marked jumps appear in Fig.~\ref{figA3}(b,c) at the times $t=t_*$ associated to the DPT, see Figs.~\ref{figA1}(b,c) and \ref{figA2}(b,c). The time dependence of $F_{\rm bw}(t)$ thus also reveals the location of the DPT, in agreement with our previous analysis.

We also define a trace distance between two correlation matrices $\theta_1$ and $\theta_2$,
\begin{equation}
{\cal D}_T [\theta_1,\theta_2]= \frac{1}{2}  {\rm Tr} |\theta_1 - \theta_2 |. 
\label{s.4.3}
\end{equation}
In principle, there is no specific relation between ${\cal D}_T [\theta_1,\theta_2]$ and the trace distance between  $\rho_1$ and $\rho_2$ \cite{Nava2024m}. However, 
from the results for ${\cal D}_T [\theta (t) ,\theta_{\rm eq}]$ in Fig.~\ref{figA3}, we observe that Eq.~\eqref{s.4.3} encompasses the key features contained in the forward and  backward fidelities. In particular, the relaxation crossover of the initially dominant harmonics of the order parameter and the location of DPTs are correctly diagnosed.  Incidentally, this observation suggests that Eq.~\eqref{s.4.3} can provide an efficient and easily computable probe for DPTs in open systems as long as the Lindblad approach is applicable.

\section{Quantum Mpemba effect} \label{appD}
   
\begin{figure}[t]
\centering  
\includegraphics[width=0.98\linewidth]{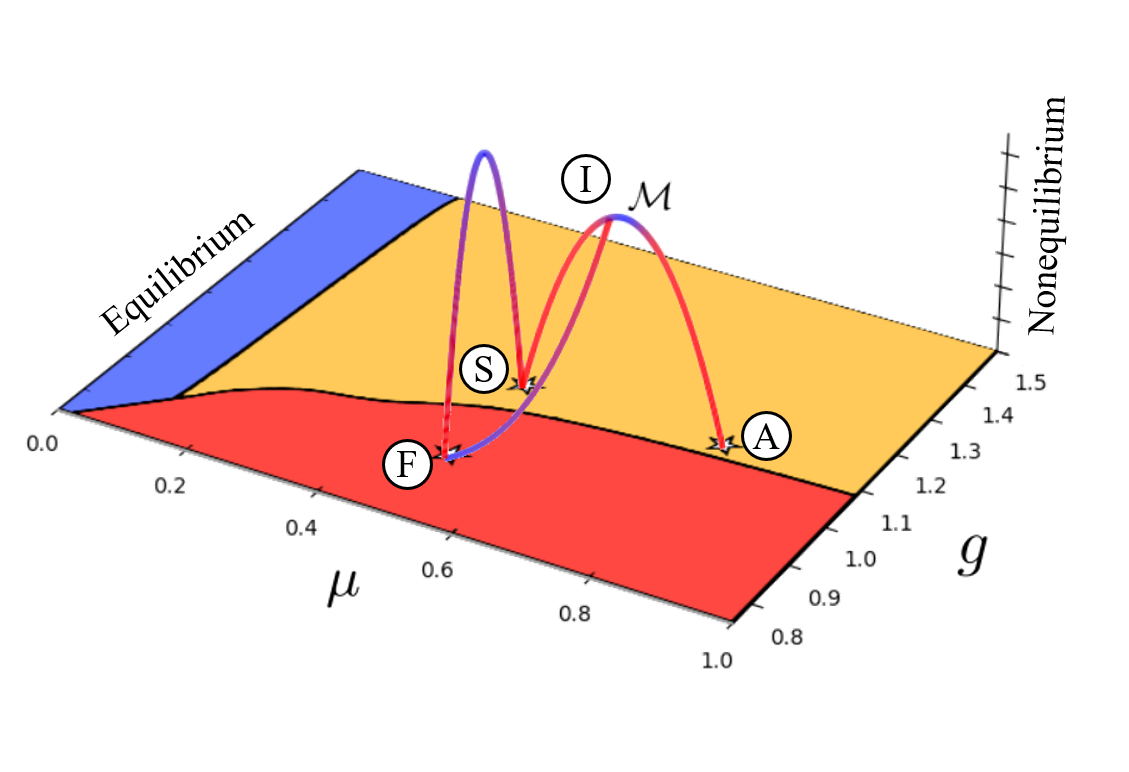}
    \caption{Schematic Pontus-Mpemba protocol for the GN model. The initial and final states ${\bf S}$ and ${\bf F}$, respectively, are thermal states and can be depicted as points in the phase diagram of Fig.~\ref{fig1}.  This is also true for the auxiliary state ${\bf A}$. However, the intermediate trajectories for both system copies in the PME refer to nonequilibrium states, including the state ${\bf I}$ where the second step of the protocol is initiated.  Since nonequilibrium states cannot be represented as points in the $\mu$--$g$ plane, the trajectories are indicated as moving out of this plane.  }    
 \label{figA4}
\end{figure}

In the main text, we have discussed the effect of DPTs on the efficiency of protocols implementing the PME.  The Pontus-Mpemba protocol is illustrated schematically in  Fig.~\ref{figA4}, where we emphasize that only the states ${\bf S}, {\bf F},$ and ${\bf A}$ are thermal states which can be represented by points in the phase diagram of Fig.~\ref{fig1}, in which the
colors of the various regions in the horizontal plane correspond to the ones of the (equilibrium) phase diagram in Fig.~\ref{fig1}.
The time evolution away from these points instead refers to nonequilibrium states, schematically indicated by moving out of the $\mu$--$g$ plane in Fig.~\ref{figA4}. For the nonequilibrium trajectories, we highlight in red (blue) color the regions characterized by a fast (slow) time evolution of the system. In this section, we address the standard single-step protocol underlying the quantum Mpemba effect (QME) \cite{Ares2025,Teza2025}, where 
one compares the dynamics for two different initial states approaching the same final state.
 
For the QME in open systems, one has to quantify the distance between the actual state and the steady state in a careful manner, both with respect to parameter space and to Hilbert space \cite{Nava2024m}.  
Denoting the set of quench parameters $p_i$ by $\{p\}$, which for our case correspond to $\mu$ and $g$, 
a parameter space distance was defined in Ref.~\cite{Nava2024m} in terms of the Euclidean distance
\begin{equation}\label{euclidean}
    {\cal D}_E [\{p_{\rm in}\},\{p_{\rm eq}\}]= 
\sqrt{\sum_i |p_{{\rm in},i} - p_{{\rm eq},i}|^2},
\end{equation} 
with pre- and post-quench parameters $\{ p_{{\rm in}} \}$ and $\{ p_{{\rm eq}} \}$, respectively.   Note that the PME does not require the introduction of a parameter distance since both system copies start from the same initial state ${\bf S}$. 
The distance between the system state $\rho(t)$ at time $t$ after the quench  and the steady state $\rho_{\rm eq}=\rho(t\to \infty)$ can be measured in terms of the trace distance \cite{Nava2024m},  
${\cal D}_T(t)= \frac{1}{2} {\rm Tr} |\rho (t) - \rho_{\rm eq} |$. 

 \begin{figure}
\centering  
\includegraphics[width=0.76\linewidth]{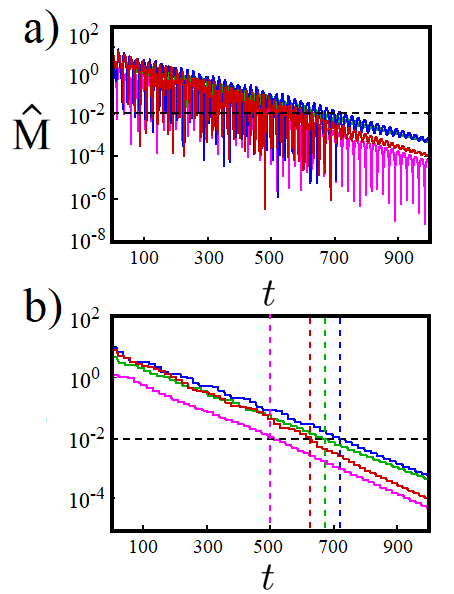}
\caption{Quantum Mpemba protocol for the GN model. (a) $\hat{M}(t)$ vs $t$ \new{(time in units of $J^{-1}$)}, see Eq.~\eqref{orderdistance},
for a 1D chain with $L=100$, $J=1$, $\gamma = 0.01$, and $k_B T=0.05$, following a quench at $t=0^+$ from  $(\mu_{{\rm in},\alpha},g_{\rm{in},\alpha})$, with $\alpha =1,2,3,4$, to  $(\mu_{\rm eq},g_{\rm eq}) = (0.5,0.9)$. Note the semi-logarithmic scales.
The black dashed horizontal line marks the threshold value $\hat{M}_*$ where we read off the
corresponding relaxation times $\tau_\alpha$.
The initial parameters are $(\mu_{{\rm in},1},g_{{\rm in},1})= (0.5,1.1)$ (green),   $(\mu_{{\rm in},2},g_{{\rm in},2})= (0.8,1.1)$ (magenta), $(\mu_{{\rm in},3},g_{{\rm in},3})= (0.5,1.3)$ (blue), and $(\mu_{{\rm in},4},g_{{\rm in},4})= (0.25,1.1)$ (red curve).
(b) Envelope functions corresponding to (a).  The relaxation times $\tau_\alpha$ for 
initial condition $\alpha$ are indicated by the vertical dashed lines; see text for a detailed  discussion.
 }    \label{figA5}
\end{figure}

To define the QME protocol, one prepares two different system copies with initial parameters 
$\{p_{\rm in}\}=\{p_{c}\}$ (``close'' to the steady state values $\{p_{\rm eq}\}$) and 
$\{p_{\rm in}\}=\{p_{f}\}$ (``far''). By definition,
 we require ${\cal D}_E [\{p_{c}\},\{p_{\rm eq}\}]
< {\cal D}_E [\{p_{f}\},\{p_{\rm eq}\}]$.  The corresponding thermal states $\rho_{c,f}$ are realized for $t<0$.  The protocol is such that both system copies, with respective initial state $\rho(0)=\rho_c$ and $\rho(0)=\rho_f$, approach $\rho_{\rm eq}$ for $t\to \infty$ --- the question is which relaxation time $\tau$ is shorter.
To that end, one compares the time evolution of $\rho_{c/f} (t)$, where the index
refers to the corresponding initial condition, after the respective 
parameter quench $\{p_{c/f}\}\to \{p_{\rm eq}\}$. 
If ${\cal D}_{T,c} (t) < {\cal D}_{T,f} (t)$ holds for all $t$, where $D_{T,c/f}(t)$ refers to the trace distance for $\rho_{c/f}(t)$,  there is no QME. At variance, a type-I QME occurs if
 ${\cal D}_{T,c } (t) >  {\cal D}_{T,f} (t)$ for all times. Finally, the most elusive
type-II QME is characterized by ${\cal D}_{T,c} (t)  >  {\cal D}_{T,f } (t)$ for times
 $t > t_*$ with a finite $t_*$; this case requires at least one crossing of the trace distance curves \cite{Nava2024m}.  

While  ${\cal D}_T(t)$ allows for an efficient detection of QMEs in small open quantum systems \cite{Nava2024m,Zatsarynna2025}, computing or measuring the trace distance is impractical or even impossible for the exponentially large Hilbert spaces of large many-body systems as encountered in our case. Moreover, since self-consistency renders the time evolution intrinsically nonlinear, additional complications arise. Therefore, while we retain the parameter distance in Eq.~\eqref{euclidean} with $\{p\}=(\mu,g)$, instead of the trace distance we here employ the order parameter distance 
 \begin{equation}\label{orderdistance}
     \hat{M} (t) = \sqrt{\sum_\nu \left [ \hat{m}(\nu,t)-\hat{m}_{\rm eq}(\nu) \right ]^2},
 \end{equation}
 which is  the non-normalized version of $M(t)$ in Eq.~\eqref{order_parameter_distance}. The reason for switching from $M(t)\to \hat M(t)$ here is that
we need to synoptically monitor time evolution patterns starting from different initial points.  
 
 To investigate the  interplay between DPTs and QME,  see Fig.~\ref{figA5},
 we have studied $\hat{M}(t)$ for four different parameter quenches. The steady state was always taken at  $(\mu_{\rm eq},g_{\rm eq}) = (0.5,0.9)$, i.e., within the DP. 
The initial parameters are all chosen to be within the CP, see the caption of Fig.~\ref{figA5},
and come with well-defined dominant harmonics $\hat{m}(\nu)$.
From Eq.~\eqref{euclidean}, we find
\begin{equation}
{\cal D}_{E,1} < {\cal D}_{E,4} <  {\cal D}_{E,2}< {\cal D}_{E,3}.
\end{equation}
The initial parameter configuration $\alpha=1$ ($\alpha=3$) is therefore closest to (farthest away from) the steady state values. Clearly, all four quench protocols take the system across two different phases, CP~$\to$~DP. Since the steady state is characterized by $\hat{M}(t \to \infty)=0$, we extract the relaxation times 
$\tau_\alpha$ for initial configuration $\alpha$ (with $\alpha=1,2,3,4$) from $\hat{M}(t)$ by setting a lower threshold, $\hat{M}(\tau_\alpha)=\hat{M}_*\sim 10^{-2}$. 
As long as $\hat{M}_*\ll 1$, the precise choice of $\hat{M}_*$ is irrelevant for the QME classification \cite{Nava2024m}. 
 
In Fig.~\ref{figA5}(a), we show the time dependence of $\hat{M}(t)$, which exhibits strong oscillations.  We note that the respective dominant momentum index $\nu$ for the pre-quench order parameter is given by $\nu=4$ for $\alpha=1$ (green curve) and $\alpha=3$ (blue), 
$\nu=7$ for $\alpha=2$ (magenta), and $\nu=2$ for $\alpha=4$  (red). 
Unfortunately, the strong oscillations in $\hat{M}(t)$ do not allow for  
 sharply identifying $\tau_\alpha$. Since $\tau_\alpha$ needs to be extracted from  
 a monotonic function of $t$, see also Ref.~\cite{Strachan2025}, we instead use 
 the upper envelope curve for each of the curves in  Fig.~\ref{figA5}(a), and show these
 envelope functions in Fig.~\ref{figA5}(b).  
 Specifically, at time $\bar{t}$, the upper envelope function is defined as the maximum value of $\hat{M}(t)$ for all times $t\geq \bar{t}$.  The resulting curves in 
 Fig.~\ref{figA5}(b) are smooth and 
 monotonic, and thus allow us to extract the relaxation times $\tau_\alpha$. 
 Apparently, we find
 \begin{equation}\label{relaxationtimes}
 \tau_3 > \tau_1 > \tau_4 > \tau_2.
 \end{equation}
As a result, we note that the relaxation time $\tau_\alpha$ depends on the distance between   $(\mu_{\rm in},g_{\rm in})$ and the phase boundary, 
rather than on  the distance between $(\mu_{\rm in},g_{\rm in})$ and $(\mu_{\rm eq}, g_{\rm eq})$. In particular, we find that a type-I QME can be realized considering the initial conditions $P_1$ (as the ``close'' one to the steady state values) and $P_2$ (as the ``far'' one from the steady state values but closer to the CP-DP phase boundary). Indeed, we have that $\tau_2<\tau_1$ and ${\cal D}_{T,1 } (t) >  {\cal D}_{T,2} (t)$ at any time. Instead, a type-II QME can be realized taking as initial conditions $P_1$ (again as the ``close'' one) and $P_4$ (as the ``far'' one from the steady state values but closer to the CP-OP phase boundary). In this case, $\tau_4<\tau_1$ and ${\cal D}_{T,1 } (0) <  {\cal D}_{T,4} (0)$ but ${\cal D}_{T,1 } (t) >  {\cal D}_{T,4} (t)$ for $t>t_*\approx500$. However, in trading the initial point $P_\alpha$ for 
a different one, $P_\beta$, closer to the phase boundary, if $P_\alpha$ and $P_\beta$ are separated by a DPT, the  intermediate region  ${\cal M}$ sets in,
see Figs.~\ref{figA1}(b) and \ref{figA2}(b) as well as Fig.~\ref{fig2}(b) for the  case where both $P_\alpha$ and $P_\beta$ lie within the CP. It turns out that the
additional time spent by the system when passing across ${\cal M}$ is much longer than the time gain from the QME, 
rendering the conventional QME useless for practical purposes. This insight is the main motivation for resorting to the PME as presented in the main text. 

\bibliography{refs}

\end{document}